\documentclass[lettersize,journal]{IEEEtran}

\usepackage{amsmath,amssymb,amsfonts}
\usepackage{algorithmic}
\usepackage{graphicx}
\usepackage{multicol}
\usepackage{multirow}
\usepackage{textcomp}
\usepackage{hyperref}
\usepackage[backend=bibtex,style=ieee,natbib=true]{biblatex} 
\def\BibTeX{{\rm B\kern-.05em{\sc i\kern-.025em b}\kern-.08em
    T\kern-.1667em\lower.7ex\hbox{E}\kern-.125emX}}
\begin{document}
\title{Recursive 3D Segmentation of Shoulder Joint with Coarse-scanned MR Image}
\author{Xiaoxiao He, Chaowei Tan, Virak Tan, and Kang Li
\thanks{This work was partially supported by National Key Research and Development Program (2020YFB1711503).}
\thanks{X. He is with Department of Computer Science, Rutgers University, Piscataway, NJ 08854 USA.}
\thanks{C. Tan is with Department of Computer Science, Rutgers University, Piscataway, NJ 08854 USA.}
\thanks{V. Tan is with Institute for Hand \& Arm Surgery, Madison, NJ 07940 USA and also a clinical professor in Rutgers-New Jersey Medical School, Newark, NJ 07103 USA.}
\thanks{K. Li is with West China Bimedical Big Data Center, Sichuan University West China Hospital, Sichuan Province, China, and also Med X Center for Infomatics, Sichuan University, Sichuan Province, China.}}

\maketitle

\begin{abstract}
For diagnosis of shoulder illness, it is essential to look at the morphology deviation of scapula and humerus from the medical images that are acquired from Magnetic Resonance (MR) imaging. However, taking high-resolution MR images is time-consuming and costly because the reduction of the physical distance between image slices causes prolonged scanning time. Moreover, due to the lack of training images, images from various sources must be utilized, which creates the issue of high variance across the dataset. Also, there are human errors among the images due to the fact that it is hard to take the spatial relationship into consideration when labeling the 3D image in low resolution. In order to combat all obstacles stated above, we develop a fully automated algorithm for segmenting the humerus and scapula bone from coarsely scanned and low-resolution MR images and a recursive learning framework that iterative utilize the generated labels for reducing the errors among segmentations and increase our dataset set for training the next round network. In this study, 50 MR images are collected from several institutions and divided into five mutually exclusive sets for carrying five-fold cross-validation. Contours that are generated by the proposed method demonstrated a high level of accuracy when compared with ground truth and the traditional method. The proposed neural network and the recursive learning scheme improve the overall quality of the segmentation on humerus and scapula on the low-resolution dataset and reduced incorrect segmentation in the ground truth, which could have a positive impact on finding the cause of shoulder pain and patient’s early relief.
\end{abstract}

\def\abstractname{Note to Practitioners}
\begin{abstract}
This paper addresses the challenge in the development of automated segmentation algorithms, which is the lacking sufficient training data and corresponding labels. Existing segmentation methods are either based on traditional methods that require hand-picked training data or machine learning-based methods that require a large number of training image-label pairs. However, acquiring both the patient data and its corresponding annotation is fairly difficult in the medical field. Therefore, we proposed a method based on recursive learning and augmentation that helps train a segmentation neural network on shoulder joint MR images better than the traditional methods with only 50 labeled images. We showed that why this method helps the network to perform better mathematically and demonstrated the superior performance of our method compared to other traditionally used methods. In future research, we will address the need to reduce the number of labels to achieve a comparable segmentation accuracy. 
\end{abstract}


\section{Introduction}
\label{sec:introduction}
\begin{figure}
    \centering
    \includegraphics[width=\columnwidth]{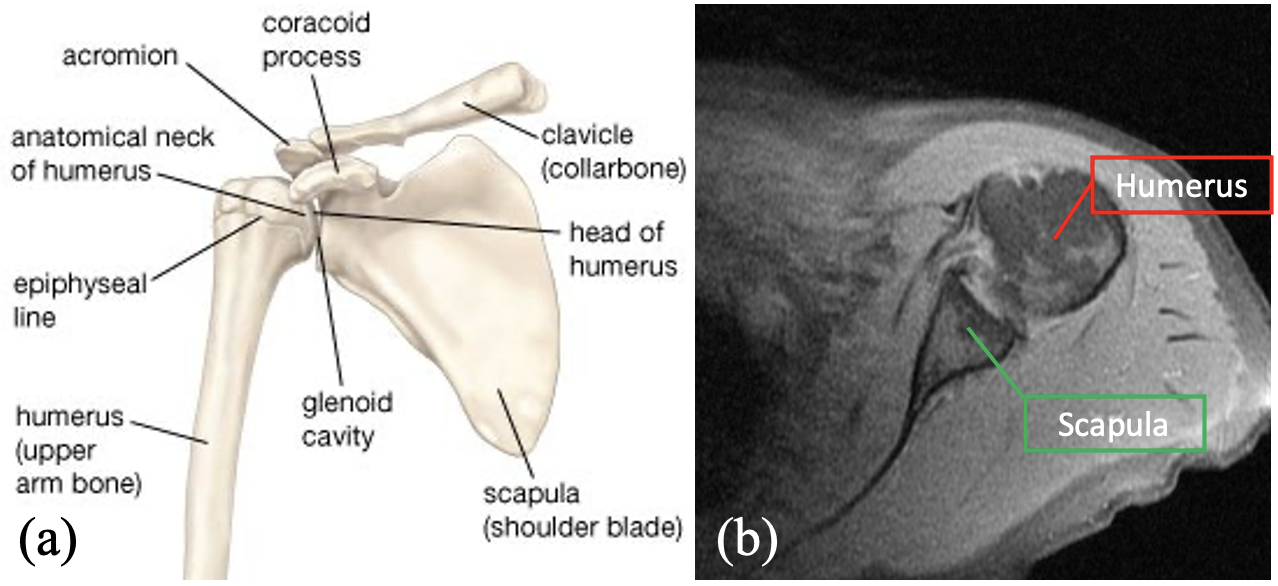}
    \caption{(a) Schematic view of shoulder anatomy (from britannica.com). (b) Demonstrative MR image slice along axial direction in our dataset.}
    \label{fig:background}
\end{figure}

\footnotetext{The source code for this work is at: \url{ https://github.com/hexiaoxiao-cs/Recursive-3D-Segmentation-of-Shoulder-Joint-with-Coarse-scanned-MR-Image}}\IEEEPARstart{S}{houlder} pain is among the most common regional pain syndromes. About $18\%-26\%$~\cite{LINAKER2015405} of the adult population had shoulder pain some time in their life. The causes of shoulder pain may involve shoulder dislocation, shoulder bone fraction, shoulder separation, shoulder arthritis, and bone tumor. Among all of those diseases, osteoarthritis is one of the common causes for shoulder pain. There are two types of osteoarthritis (OA): primary OA and secondary OA. Primary OA is caused by normal wear and tear of the shoulder joint, which is a common cause of shoulder pain among patients who are above 60 years old. Secondary OA requires the presence of other diseases or conditions. When a patient is diagnosed with OA, the doctor may suggest one of the following three operations, hemiarthroplasty, total shoulder arthroplasty (TSA), or reverse total shoulder arthroplasty (RTSA). Hemiarthroplasty is preferred in the presence of fractures or avascular necrosis in the humeral head is inevitable \cite{Mattei2015}, while TSA or RTSA is performed when the patient is diagnosed with primary OA or secondary OA, respectively. Preoperative diagnosis that is based on studying the morphology of human humerus and scapula from the 3D medical images including computed tomography (CT) and magnetic resonance (MR) imaging is needed for all three operations, which includes multiple analysis and measuring tasks on assessing shoulder instability. Because of the radiation absorption during a CT scan, the use of CT imaging technology is limited. Therefore, MR imaging has a greater reach. A typical 3D model of a humerus and scapula bone system is shown in Fig.~\ref{fig:background} (a). In the 3D image, we can see the complexity of the humeroscapular interface, and the connecting area of both bones, which is the glenoid cavity, may go through extensive wear and tear, thus can cause deformation. Therefore, the segmentation of such areas is vital to clinical diagnostic. Nevertheless, the structure of the scapula bone creates a barrier for segmentation tasks because of the following two reasons. First, the coracoid process and acromion are close to the humerus bone since the functionality of such two bones is for the stability of the shoulder joint. Thus, it is hard to separate them apart. Second, the concaveness of the glenoid cavity creates difficulties for the traditional segmentation method \cite{1490669}. Usually, the diagnosis is made by a doctor reading a series of 2D medical images one after another. Such process requires the doctor to interpret the spatial relationship between 2D slices, which is a hard job to do. Instead, we can utilize a volumetric MR image and perform segmentation on that 3D image in order to get a more straightforward visualization of the patient's shoulder area.

\begin{figure}
    \centering    \includegraphics[width=\columnwidth]{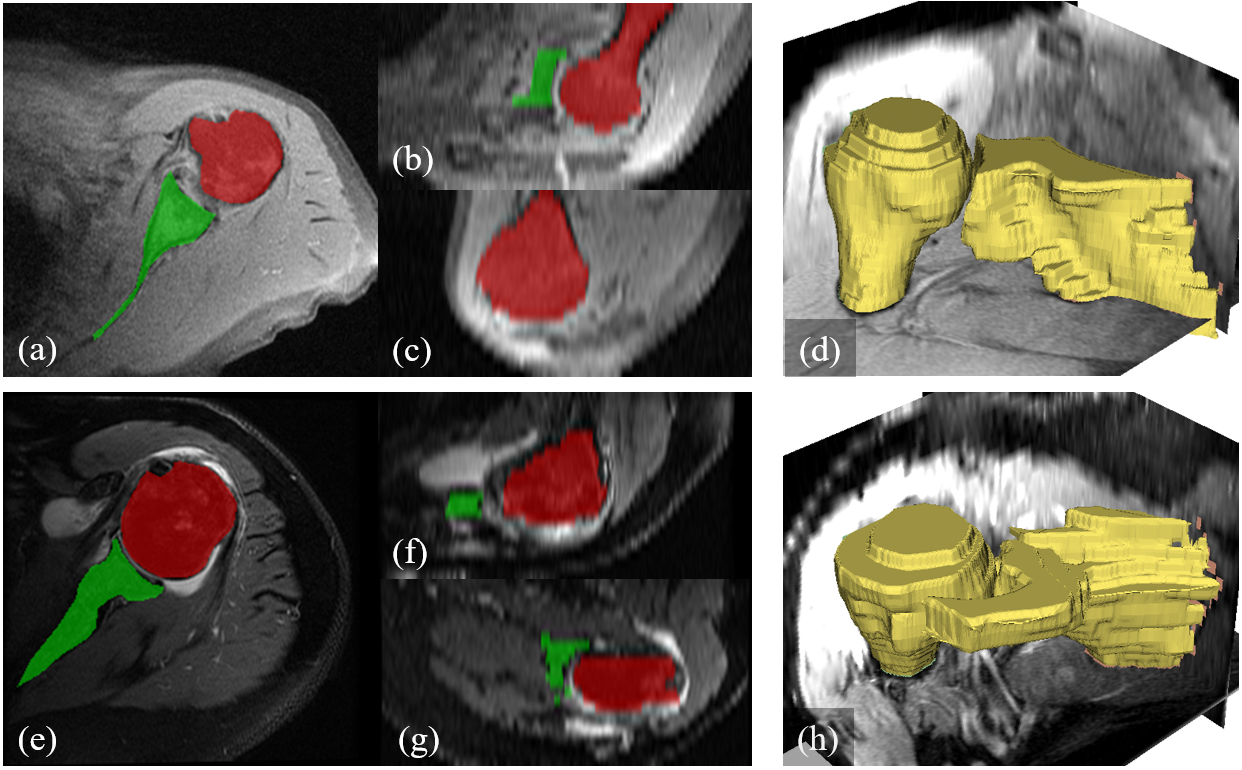}
    \caption{(a)-(c) and (e)-(g) represent  axial, coronal, sagittal slices of two 3D MR shoulder data respectively. (d) and (h) demonstrate the humerus (left bone) and scapula (right bone) label in 3D.}
    \label{fig:problem_statement}
\end{figure}

2D MR images stacked along a specific scanning direction build the volumetric image. There are works done by using high-resolution MR images \cite{9098360} which take advantage of reduced spacing between 2D slices in order to capture more details. Although a high-resolution image may be beneficial for the diagnostic, it is hardly a practical choice since it prolongs the scanning time and may cause the following challenges: First, the patient needs to remain still for an extended period of time in the machine. Otherwise, there will be blurred areas in the scan, which undermines the quality of the image. Second, the demand for such diagnostic a tool is high, and the increasing imaging time can reduce the number of people get treated every day. Moreover, the prolonged imaging time will create a financial burden to the patient, which may limit the acceptability of such a tool. Therefore, coarse scanned MR image is more common in the clinic compared to the high-resolution MR image. Fig.~\ref{fig:problem_statement} demonstrates our image dataset, which contains 50 coarse scanned Mr images with spacing ranging from $1.0$mm to $4.5$mm along the axial axis. Fig.~\ref{fig:problem_statement} (b), (c) and (f), (g) corresponds to coronal and sagittal slices respectively, which resolutions are much lower compared to Fig.~\ref{fig:problem_statement} (a) and (e). This is because our coarse scanned MR images are clinical data and thus constrained by cost and time. Moreover, comparing Fig.~\ref{fig:problem_statement} (a) and (e), the relative image contrast between the bone and the surrounding muscles is different since our image comes from various sources. Such low resolution will create layer-liked artifacts as demonstrated in Fig.~\ref{fig:problem_statement} (d) and (h) because of the loss of details between image slices. Furthermore, because our dataset contains images captured with various configurations from different imaging facilities, the parameters of images, such as contrast and size, differ from each other. In order to extract useful volumetric information from the 3D MR images, separating and labeling the humerus and scapula from other tissues is required for the diagnostic of bone illness. The manual marking on a 3D image is usually done by labeling multiple 2D images, which makes it a tedious, laborious, and time-consuming task, especially if the size of the image is large and the spacing is small. Moreover, the task of manual segmenting 3D images is hard because the lack of spatial sense during labeling one 2D image, and thus may create discontinuity in the surface mesh generated by the masks. Therefore, it is necessary to develop a 3D segmentation algorithm that is reliable and effective and can offer repeatability in different medical applications.

For bone segmentation in Magnetic Renascence imaging, there are well-established methods involving human interaction. T. Stammberger et al. proposed an active contour segmentation method utilizing B-spine snake algorithm \cite{STAMMBERGER19991033}, which takes human input as seed input and utilizes b-spline algorithm for interpolation. Although the methods increase the efficiency of labor compared with total manual segmentation, it still requires some degree of human input. N.T. Nguyen et al. \cite{4228556} showed that region growing methods provide good segmentation results on MR images with sharp edges and reduce the degree of human intervention, but the method relies heavily on the quality of the image and is not computational efficient in the process of finding similar pixels and may fail to provide a success segmentation due to noises in the image. Moreover, most of the traditional methods \cite{STAMMBERGER19991033,4228556} are 2D-based segmentation methods, which may cause the surface mesh to exhibit strong spatially unrelated characteristics like the sliced surface. Some of the methods utilize MR and CT image from the same patient in order to perform fully automatic segmentation. J. Folkesson et al. \cite{doi:10.1118/1.3264615} proposed a binary approximated kNN classifier for 3D image segmentation. The method utilizes fussy C-means clustering with local second-order features for bone enhancement. Yet, the method yields a good result with low-resolution MR images; the segmentation process is relatively slow, and the time it needs proliferates with the increasing amount of clusters. Another research by S.P. Raya et al. introduced a rule-based segmentation method\cite{57771} which utilize a predefined database of rules to perform segmentation on brain MR image. The rules are based on the histogram of low-level features in order to determine the likelihood that the pixel is in the class. Due to the overlapping of the histogram from several tissues, Raya et al. used T1 and PD weighted MR images to solve such problems. Therefore, it is evident that the rule-based system can be both computation-consuming and erroneous. 

Another important method for MR image segmentation is based on registration. An atlas is a pair of image scan with the corresponding manual labeling mask.  Wang et al. \cite{wang2013multi} proposed an atlas-based segmentation method that compares similarity within atlases and between atlases and the target image in order to yield a high-quality label. The method computes the voting weights by minimizing the fusion error. However, the accuracy of such method is strongly related to the sparseness of the distribution and quality of the atlases since the success of performing segmentation needs the target image and atlas to be close.  Another drawback for multi-atlas with joint label fusion method is that it requires a set of expert-selected atlases so that the atlas dataset can contain as many features as possible. The deformable shape is another registration-based segmentation method. G.B. Sharma et al. \cite{sharma2013adaptive} introduced a segmentation workflow that generates a deformable shape model by using real bone CT scans. Instead of using a single fixed bone for building the global SSM, T. Mutsvangwa et al. \cite{mutsvangwa2014an} provided us an alternative way of producing labels based on medical images by employing an iterative median closest point-Gaussian mixture model (IMCP-GMM) method. A different SSM approach \cite{SCHMID2011155} that utilizes a small field of view of the MR images and multi-resolution algorithm with adaptive initialization yields reputable results. Nevertheless, all three statistic shape models are reasonably accurate since they are based on an actual bone or a consensus of MR images. The ability of generalization remains to be questioned. 
\begin{figure*}
    \begin{center}
    \begin{tabular}{c} 
        \includegraphics[height=0.26\textheight]{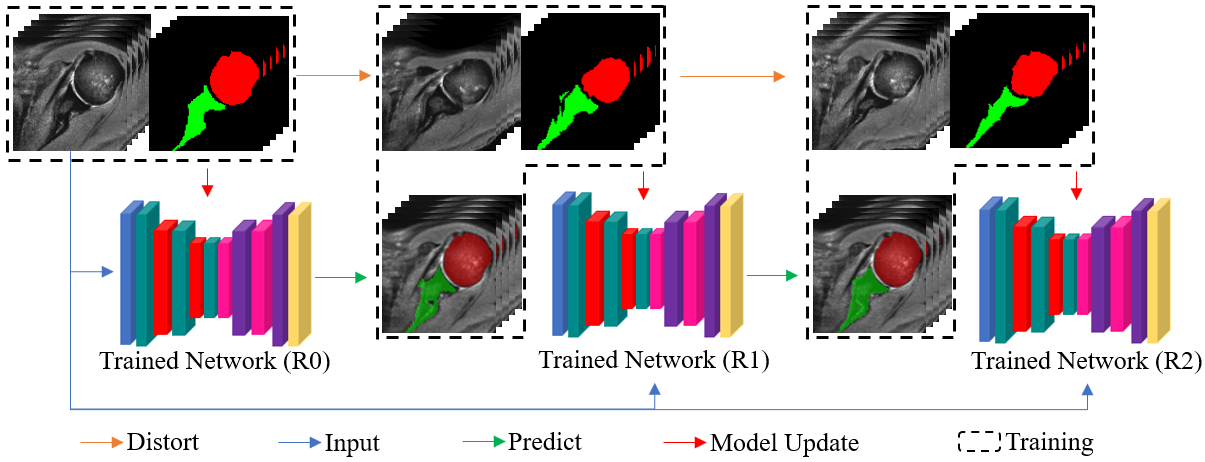}
    \end{tabular}
    \caption[example]{\label{fig:abstrain} Flow chart of the proposed recursive learning method. After the initial segmentation network (R0), the trained network from previous round is used to segment the input images in order to get labels for the current round training. Then we distort the initial training image and ground truth. By utilize the distorted training pairs and undistorted generated pairs, we train the current segmentation network. Repeat these steps until we reach the third neural network.}
    \end{center}
\end{figure*}

Benefit from the recent development of deep learning that utilizes convolutional neural network to produce labels, several automatic and semi-automatic approaches are developed to reduce the amount of human labor due to the high cost of human labels. However, the majority of the researchers focus on the segmentation of fine scanned CT shoulder bones images \cite{sharma2013adaptive}. Although the CT scanning could offer high-contrast bone imaging, there are still concerns about its moderate to high radiation and thus may lead to limitation on wide application for high-resolution and large-region 3D shoulder imaging. On the other hand, MR imaging can provide direct and noninvasive images of bone joints. However, because the demand for MR imaging is high, obtaining high-resolution (i.e., low slice thickness) MR images is quite costly and time-consuming clinically as stated previously. In addition, the parameters of scanning may differ from case to case, which will increase the variance of images and creates difficulty for clinical use. Therefore, the healthcare industry has a need for a fully automated method for 3D shoulder joint segmentation that is designed for MR images with the ability to perform successful segmentation on a coarse scanned image with high variance. In order to work out all the difficulties listed above, we proposed a recursive learning framework which is demonstrated in Fig.~\ref{fig:abstrain}. In the proposed framework, we divide the whole training process into three rounds. For each round, the training images together with their labels will be used to train the current network, and then the trained network will be used to generate the labels based on the input images of the current round. The following segmentation network will be trained with data combining the input images and the corresponding generated label with the distorted input images and labels. Because the initial training images and labels are limited, such self-generating training data will help the next round to achieve better performance. Moreover, the proposed method has the ability to correct some of the mistakes made in the ground truth because the generated labels represent features that are more common among our training dataset, and by training with more generated labels, the less likely that errors in ground truth would reappear.

In this work, we propose a novel recursive learning framework together with a deep end-to-end segmentation network that focuses on segmenting a small MR image dataset which contains low-resolution images. For solving problems brought by such great variance and inconsistency among images while having a limited amount of data (50 images in our experiment), we propose the recursive learning framework. In the proposed framework, we utilize data augmentation technique to increase the training set size for reducing human errors. We adopt the recent U-net network architecture \cite{isensee2017brain} and propose a deep neural network that reduces the number of layers and thus the required training data size. Our major contributions are mainly two parts: 1. The task of 3D segmentation with low resolution, limited amount of data and high shape variability is addressed for the first time. We demonstrate that the problem of limited data could be dealt with data augmentation produced in a recurrence fashion. 2. We propose a streamlined recursive learning framework that is designed to help underlying neural network better training with minimal data. 

\section{Related Work}
\label{sec:ReL}
\subsection{Semi-automatic Segmentation} Traditionally, the semi-automatic segmentation methods include empirical threshold approach, seeded region growing, active contour model, and cluster-based segmentation methods. The commonly used segmentation method for osteoporosis-related studies is setting a threshold interval where all pixels with intensity lying in the interval will be labeled. S. Majumdar et al. \cite{jbmr.1997.12.1.111} use the intensity histogram of a selected region of interest for determining the threshold. Such method does not produce a fairly accurate segmentation, especially on the edge of the target bone, since it does not take nearby pixels into consideration during segmentation and thus would reduce the accuracy. Ryba et al. \cite{ryba2019segmentation} proposed a semi-automatic segmentation of MR image of the shoulder joint utilizing watershed-based region growing for coarse segmentation and active contour model for refining produced label. Although their method showed promising results for segmenting humerus on a 2D MR image, whether the method could apply to a 3D problem remains questioned. Nguyen et al. treated the task as a boundary-based segmentation problem and utilized region growing methods that require the user to input at least one point. This is significantly less human work compared to the previous method, but the iterating nature of the method made the segmentation process inefficient. J. Folkesson et al. treated the segmentation as a voxel-based clustering problem and utilized fuzzy-probabilistic C-means and shape-based topological property in order to create a segmentation of brain tumors.

\subsection{Fully Automated Segmentation} In order to reduce the human labor with the rising demand for medical imaging, a fully automated segmentation method is needed. Approaches which do not need human interference including atlas-based registration and statistical shape models (SSMs) of bony structures constitute the foundations for segmentation of shoulder joint on MR images. By registering the target image to the atlases, a label mask is predicted, and hence the target image is propagated to single or multiple atlases to extract and fuse the final segmented masks. Wang et al. proposed a representative multi-atlas approach by using joint label fusion \cite{wang2013multi}. With minimizing a total fusion error by computing the voting weights of the atlas set, such method achieves good accuracy. However, such method relies heavily on the selection of the atlas images, which limits the generalization ability with regard to unforeseen shoulder joint illness. SSM-based segmentation method can be viewed as an atlas-based method but with a different approach. Instead of merely registering the target image from the atlases, SSM will try to find a distribution of shapes from the given training set, which improves the generalization ability of the method compared to the previous method. Inspired by the iterative medican closest point-Gaussian mixture model (IMCP-GMM), an improved pipeline for creating automated unbiased global SSMs is proposed by Mutsvangwa et al \cite{mutsvangwa2014an}. Compared to the semi-automatic segmentation method introduced above, both altas-based method and SSM method perform well with difference in shape and morphology between target image and atlases. However, they still require certain amount of hand-picked training images and the proper choice of such training set in order to compute the label fusion weights or create mean shape. Without relying upon registration, which would fail if the target image is far from any images in the atlas dataset, our method learns the local features of the image together with the global features to produce an accurate segmentation. Such features do not change a lot across images, which makes our methods not severely impacted by the choice of training dataset. Furthermore, with the complicated imaging conditions mentioned above, the accuracy of labels generated by both atlas-based method and SSMs will be impacted negatively because of the non-continuous surface derived from the manually labeled ground truth. Our method utilizes 3D convolutional blocks, which inherently takes the relationship between slices in all directions into consideration, thus produces a smoother surface. 

For combating the difficulties, including limited generalize ability and the reliance on the quality of the picked atlas dataset, deep learning architectures are introduced into the field of medical image segmentation, e.g.,  UNet~\cite{10.1007/978-3-319-24574-4_28} and Hyper Dense Net~\cite{8515234}, because of their ability to extract both local and global features. Since features are an abstraction of relationship among pixels, they tend to perform and generalize better compared to registration based methods. The original UNet treated the 3D volumetric data as a series of 2D slices and performed segmentation on each slice. The results were, although fairly accurate if only viewed in 2D slices, discontinuous when we stack the 2D labels to 3D volume and generate surface mesh based on the volume. Therefore, we design our deep end-to-end network to utilize 3D convolutional blocks, which are optimal for 3D medical applications. 

\subsection{Data Augmentation} Training these 3D models requires a large amount of 3D data (over 100 subjects). In the medical field, especially for the applications about humerus and scapula, the availability of data is a great challenge.  Hemke et al. \cite{hemke2020deep} incorporated random rotation, flipping, cropping, and scaling before training their neural network. The accuracy of segmentation gained as the result of increasing training data. Compared to our training dataset, Hemke et al. already had 200 images without augmentation, which is sufficient to train a high-quality segmentation network. Moreover, the augmented image shared the same distribution as the original image since the data augmentation technique applied is pretty simple. Therefore, we proposed our recursive learning framework that applied b-spline interpolation as data augmentation in vivo. Unlike previous image augmentation that changes the image as a whole, our method creates detailed local variation that could benefit the network to learn the localized features instead of memorizing the location. Moreover, because our method use generated data as another source of augmentation, errors inhibited among ground truth can be reduced since they are unique features, and by introducing labels that contain more generalized information, we can reduce the portion of such error among our training data.
\begin{figure*}[t]
    \centering
    \includegraphics[height=0.24\textheight]{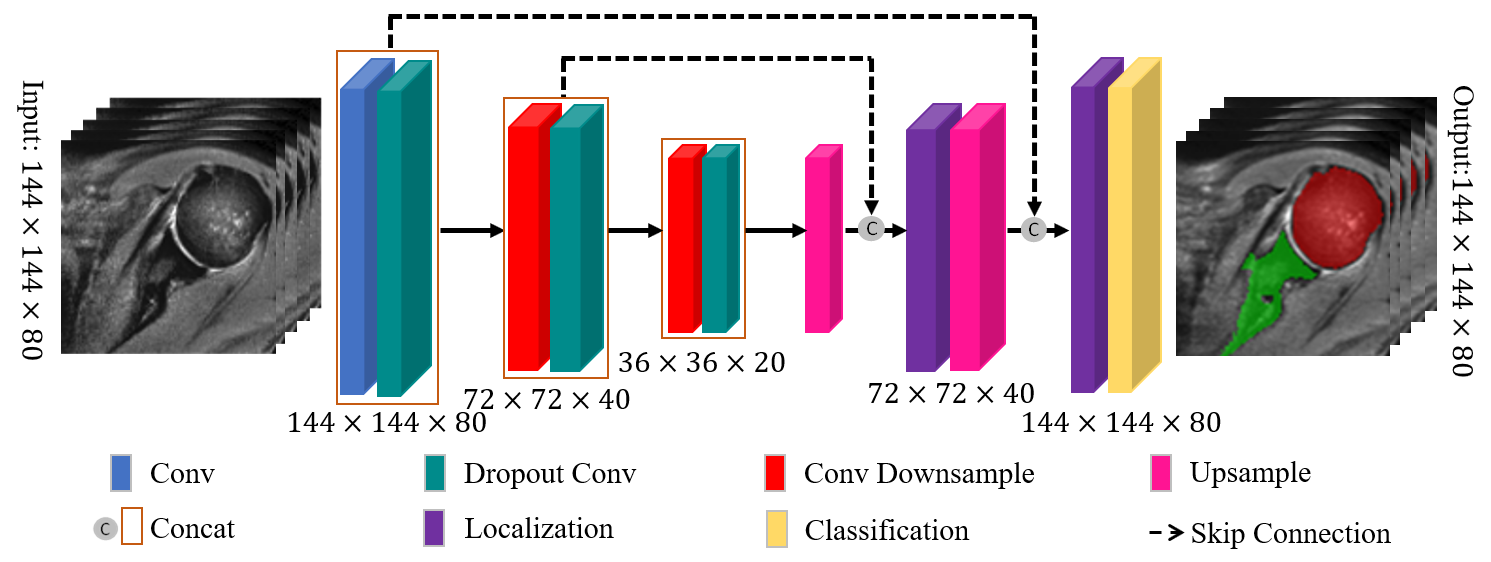}
\caption{The proposed segmentation network structure.}
\label{fig:NetworkStructure} 
\end{figure*}

\section{Methodology}
\label{sec:MaM}

The architecture of the proposed method is illustrated in Fig.~\ref{fig:abstrain}, which contains two parts: the deep end-to-end segmentation network and the recursive learning framework. The neural network primarily focuses on the segmentation task, while the recursive learning framework improves the segmentation network performance by generating and incorporating more training data for training the next-level neural network. 

\subsection{Segmentation Model}
\label{sec:DE2ENN}

Here in Fig.~\ref{fig:NetworkStructure}, our proposed segmentation network utilize a symmetric structure with an encoding path and a decoding path. The network learns a way that it could encode the preprocessed image section into features in the latent space, which are used to describe the shoulder joint abstractly. Then the network decodes the latent features into the corresponding labels. The abstraction of features in the MR image ensures the robustness of the trained model when dealing with new patients with various deformations in both scapula and humerus. Because of the high graphic card memory consumption brought by using our 3D MR shoulder data to train the original 3D UNet architecture ~\cite{10.1007/978-3-319-24574-4_28}, we made some modifications on the network architecture including reducing the depth of both encoder and decoder together with the corresponding pooling and downsampling operations to 2, respectively. At each downsampling level along the encoding direction, we use a convolution block with residual connection in order to extract feature. Here we utilize the dropout block between two consequential convolutional blocks for improving the robustness of our neural network because our lack of data requires the network to produce label based on a more representative feature and reduce the chances for overfitting. In order to help the network to acquire the integral bone structure together with the background knowledge around the target tissue, we utilize the multi-resolution encoder structure for obtaining the multi-size contextual information. In the decoding part, we first upsample the input in order to achieve the size of the previous layer by copying the image twice in each direction and then followed by a convolutional block to reduce the number of feature maps by half. Then after the skip connection that concatenates the up-sampled low-level feature maps with the skipped equivalent-resolution maps from the encoding half, we utilize a localization block~\cite{isensee2017brain} to fuse the concatenated features. A softmax activation layer is applied to the output of the final localization block in order to get a normalized probability of the pixel belongs to the corresponding class. 
Define $\phi_{j}^{i}$ as the network for $i$-th iteration on $j$-th image, and $x_j$, $y_j$ be the training image and ground truth respectively. Then the loss function for initial training can be defined by utilizing the multi-class cross entropy, 
\begin{align}
    \mathcal{L}(x,y)=-\Sigma_{j=0}^{C-1}y_j ln(\phi_{j}^{0}(x_j))
\end{align} Here $C=3$ is the number of classes, with $x_j\in \{0,1,2\}$ representing the class of background, humerus and scapula  respectively. As will be shown in the later sections, the segmentation network can produce reasonable segmentation, but can be improved by the recursive learning framework.
\subsection{Recursive Learning}
Although the segmentation network in section \ref{sec:DE2ENN} could produce average bone labels based on our flawed GT dataset and scarce training data, the segmentation accuracy of the network still can be improved. We identified several drawbacks in using only the neural network. With few and low-quality data, some of the parameters in the UNet structure would not be optimal, which may yield erroneous segmentations. Moreover, by only using a limited amount of images, the generalizing ability of the model could be questioned when segmenting a MR image with new diseases. Since there exist errors among the ground truth, we need to design a learning framework that somehow reduces such mistakes. It comes to our minds that Long Short Term Memory (or LSTM) neural networks have the ability to forget one or more pieces of information by utilizing a forget gate. LSTM is one kind of recurrence neural network that specialized in processing data that have sequential and contextual properties, which means the sequence of data and the interrelationship between data points are crucial for completing the task. LSTM takes both properties into consideration by utilizing the recurrent network structure. Moreover, the introduction of the forget gate allows the network to process a larger quantity of inputs while selecting the relevant temporal features. Because of its ability to pass down the learned features from the previous network with selection, we proposed a recursive learning strategy, and the learning process is demonstrated in Fig.~\ref{fig:abstrain}. After finishing the initial training of the network, the trained model is utilized to generate higher quality labels, and some augmentation techniques (e.g., distortion) are combined to extend the training set. Thus, we could apply the intermediate results for the subsequent training rounds. With image augmentation, images with different morphology are created to insure the robustness of the network when dealing with a new patient.

\begin{figure*}
    \centering
    \includegraphics[width=0.67\linewidth]{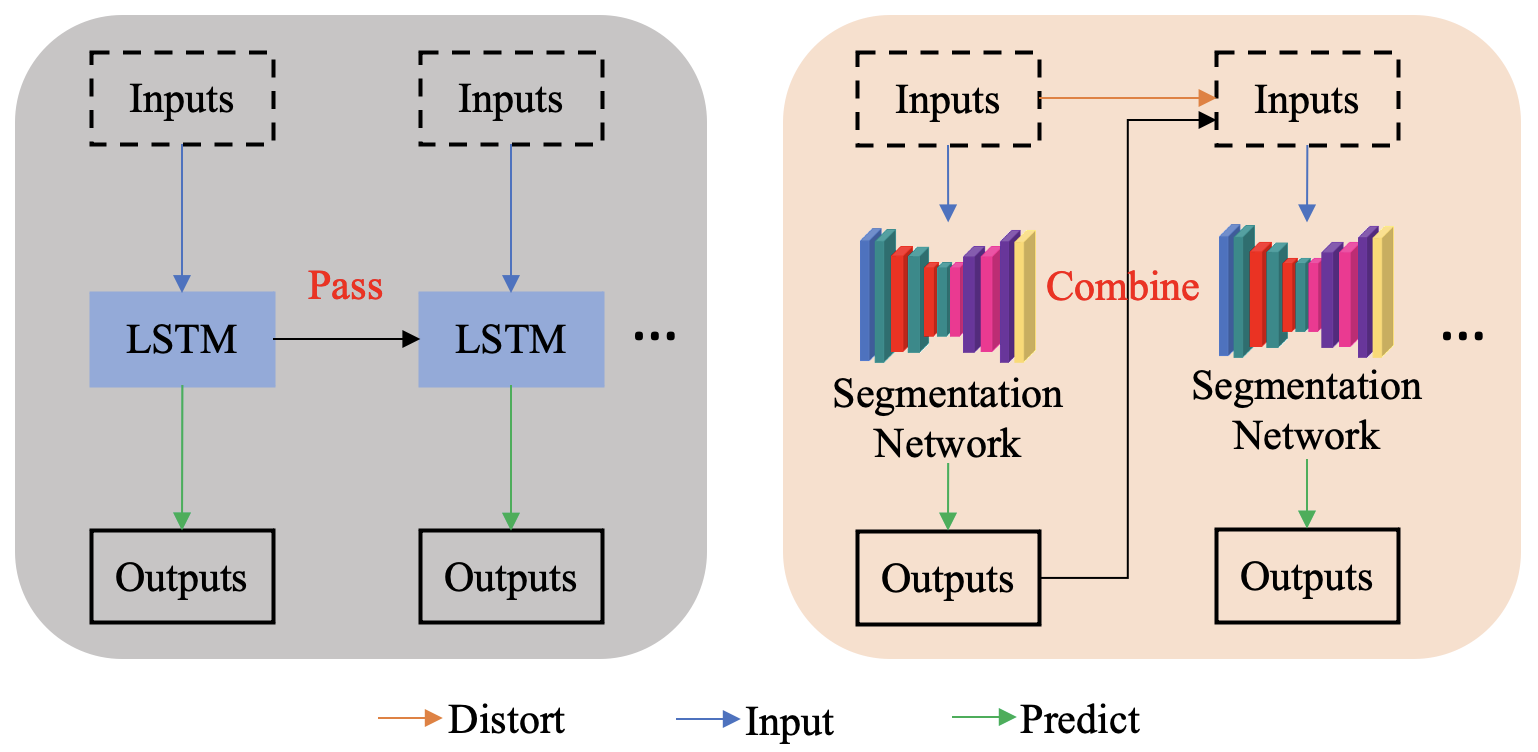}
    \caption{Left is an abstract architecture for LSTM as a recurrent neural network. Right is the proposed recursive learning framework.}
    \label{fig:method_lstm}
\end{figure*}

Here, we will perform a side-by-side analysis on LSTM and our recursive learning framework that is inspired by it. A typical LSTM structure, as shown in Fig.~\ref{fig:method_lstm} inherits the previous states by passing down features, which makes it a better choice for processing inputs that have interrelationships such as time series or words in the sentences, where training on the same structure benefits such scenario. In our case, our MR images come from different patients and different imaging facilities, which does not share the same strong connection as previous examples. Instead, we pass down the important features as the previous network learned to the next level network and reinforce such features. As for forgetting the erroneous labels among the ground truth, we can dilute the features extracted from those wrong labels by adding more common features and reducing their chances of appearance since errors are not the majority of our ground truth. Another benefit we gained from such training scheme is that by not training the same network for the next round, we have a lower chance of sinking into a local minimal since the network is trained from the fresh start. Here, we will give a more detailed analysis of the reason such framework will improve the original segmentation result. Consider the loss function for the segmentation network:
\begin{equation}
\mathcal{L}(x,y)=-\Sigma_{j=0}^{C-1}y_j ln(\phi_{j}^{0}(x_j))
\end{equation}
Then after the first training, we distort the training label and image together and add the predicted label generated with the original training image into the next training iteration. The loss for the second iteration is \begin{align}
\mathcal{L}(\hat{x},\hat{y})&=-\Sigma_{j=0}^{C-1}\hat{y_j}ln(\phi_j^{1}(\hat{x})) \\
\mathcal{L}(x,\sigma(\phi^0(x)))&=-\Sigma_{j=0}^{C-1}\sigma(\phi^0(x_j))ln(\phi_j^{1}(x))
\end{align}
Since the distortion we made does not change the characteristics of the original image, we assume that $x\approx\hat{x}$ and $y\approx\hat{y}$, then we have:

\begin{align}
     \mathcal{L}(\hat{x},\hat{y})&+\mathcal{L}(x,\sigma(\phi^0(x)))\\& = -\Sigma_{j=0}^{C-1}\hat{y_j}ln(\phi_j^{1}(\hat{x})) +-\Sigma_{j=0}^{C-1}\sigma(\phi^0(x_j))ln(\phi_j^{1}(x)) \\
    & \approx -\Sigma_{j=0}^{C-1}y_jln(\phi_j^{1}(x))
    +-\Sigma_{j=0}^{C-1}\sigma(\phi^0(x_j))ln(\phi_j^{1}(x))\\
    & = -\Sigma_{j=0}^{C-1}(y+\sigma(\phi^0(x)))ln(\phi_j^1(x)) \\
    & = \mathcal{L}(x,(y+\sigma(\phi^0(x))))
\end{align}

Here $\sigma(\phi^0(x))$ is the segmentation label produced by learning the features of our initial training dataset. Thus it represents a more common feature/pattern extraction. Since the human error does exist in our dataset but in a small amount, such errors may be reduced or eliminated due to not being a majority in the dataset. The term $\sigma(\phi^0(x))+y$ gives more weights on more common features, but also includes features that haven't been learned in the first network but are potentially useful since we preserve the original data through concatenation. Moreover, such training framework will force the network to focus on the local patterns instead of memorizing the shape, which eliminates the drawback of the registration-based methods. Furthermore, the 3D surface meshes derived from the produced segmentation labels are more continuous and accurate. Although we can perform recursive learning endlessly, empirically, two extra training rounds in our recursive learning framework could help the model to reach the highest performance and converge. This process reduces the amount of raw MR images we need and the human labor for segmenting those images.
\section{Experiment}

\subsection{Patient Data}
We acquired 50 coarse scanned MR images intended for fast localization. The size of the image ranges from $208\times 208\times 22$ to $512\times 512\times 40$ and the resolution varies from $0.3125 \times 0.3125\times 1.0$ mm to $0.9615\times 0.9615 \times 4.5$mm. Because of the limited amount of available data we have, we randomly split the $50$ images into $5$ lists and train the network with $4$ lists and validate the network using the remaining one. 
\begin{table*}[ht!]
    \centering
    \caption{Quantitative comparisons for the recursive learning using 5-fold cross-validation}
    \begin{tabular}{|l|c|c|c|c|c|c|c|c|c|}
        \hline
        \multirow{2}{*}{\textbf{\textit{G1}}} & \multicolumn{3}{c|}{\textbf{Humerus}} & \multicolumn{3}{c|}{\textbf{Scapula}} & \multicolumn{3}{c|}{\textbf{Both}}\\
        \cline{2-10}
        & \textit{DSC} & \textit{HD} & \textit{ASD} & \textit{DSC} & \textit{HD} & \textit{ASD} & \textit{DSC} & \textit{HD} & \textit{ASD} \\
        \hline
        R0 & 0.88 & \textbf{8.03} & 1.29 & 0.69 & 26.69 & 2.14 & 0.81 & 25.47 & 1.73\\
        \hline
        R1 & 0.88 & 8.20 & 1.18 & 0.66 & 25.96 & 1.82 & 0.81 & 25.10 & 1.52\\
        \hline
        R2 & \textbf{0.89} & 8.44 & \textbf{1.12} & \textbf{0.72} & \textbf{21.40} & \textbf{1.58} & \textbf{0.83} & \textbf{18.88} & \textbf{1.36}\\
        \hline
        \hline
        \multirow{2}{*}{\textbf{\textit{G2}}} & \multicolumn{3}{c|}{\textbf{Humerus}} & \multicolumn{3}{c|}{\textbf{Scapula}} & \multicolumn{3}{c|}{\textbf{Both}}\\
        \cline{2-10}
        & \textit{DSC} & \textit{HD} & \textit{ASD} & \textit{DSC} & \textit{HD} & \textit{ASD} & \textit{DSC} & \textit{HD} & \textit{ASD} \\
        \hline
        R0 & 0.87 & 8.58 & 1.33 & 0.61 & 31.18 & 1.59 & 0.78 & 29.64 & 1.51\\
        \hline
        R1 & \textbf{0.88} & \textbf{6.72} & \textbf{1.11} & 0.62 & 29.45 & \textbf{1.19} & 0.80 & 28.56 & \textbf{1.21}\\
        \hline
        R2 & 0.87 & 8.65 & 1.13 & \textbf{0.67} & \textbf{27.02} & 1.50 & \textbf{0.80} & \textbf{24.77} & 1.32\\
        \hline
         \hline
        \multirow{2}{*}{\textbf{\textit{G3}}} & \multicolumn{3}{c|}{\textbf{Humerus}} & \multicolumn{3}{c|}{\textbf{Scapula}} & \multicolumn{3}{c|}{\textbf{Both}}\\
        \cline{2-10}
        & \textit{DSC} & \textit{HD} & \textit{ASD} & \textit{DSC} & \textit{HD} & \textit{ASD} & \textit{DSC} & \textit{HD} & \textit{ASD} \\
        \hline
        R0 & 0.78 & 12.87 & 2.16 & 0.39 & 40.14 & 3.32 & 0.66 & 32.80 & 2.20\\
        \hline
        R1 & 0.83 & 12.28 & 1.66 & 0.55 & 26.99 & \textbf{2.26} & 0.75 & 24.02 & 1.79\\
        \hline
        R2 & \textbf{0.88} & \textbf{10.40} & \textbf{1.20} & \textbf{0.61} & \textbf{19.20} & 2.96 & \textbf{0.80} & \textbf{18.74} & \textbf{1.60}\\
        \hline
        \hline
        \multirow{2}{*}{\textbf{\textit{G4}}} & \multicolumn{3}{c|}{\textbf{Humerus}} & \multicolumn{3}{c|}{\textbf{Scapula}} & \multicolumn{3}{c|}{\textbf{Both}}\\
        \cline{2-10}
        & \textit{DSC} & \textit{HD} & \textit{ASD} & \textit{DSC} & \textit{HD} & \textit{ASD} & \textit{DSC} & \textit{HD} & \textit{ASD} \\
        \hline
        R0 & 0.79 & 13.85 & 2.39 & 0.49 & 34.39 & 1.39 & 0.69 & 32.44 & 2.00\\
        \hline
        R1 & 0.81 & 15.82 & 2.57 & \textbf{0.58} & 34.75 & 1.37 & 0.73 & 33.83 & 2.00\\
        \hline
        R2 & \textbf{0.86} & \textbf{10.50} & \textbf{1.48} & 0.56 & \textbf{35.49} & \textbf{1.08} & \textbf{0.76} & \textbf{30.23} & \textbf{1.33}\\
        \hline
        \hline
        \multirow{2}{*}{\textbf{\textit{G5}}} & \multicolumn{3}{c|}{\textbf{Humerus}} & \multicolumn{3}{c|}{\textbf{Scapula}} & \multicolumn{3}{c|}{\textbf{Both}}\\
        \cline{2-10}
        & \textit{DSC} & \textit{HD} & \textit{ASD} & \textit{DSC} & \textit{HD} & \textit{ASD} & \textit{DSC} & \textit{HD} & \textit{ASD} \\
        \hline
        R0 & 0.87 & 12.35 & 1.93 & 0.67 & 25.34 & 2.36 & 0.80 & 25.37 & 2.00\\
        \hline
        R1 & 0.87 & \textbf{8.61} & 1.26 & 0.65 & 26.48 & \textbf{1.47} & 0.80 & 25.96 & 1.37\\
        \hline
        R2 & \textbf{0.88} & 10.54 & \textbf{1.19} & \textbf{0.72} & \textbf{22.23} & 1.47 & \textbf{0.82} & \textbf{22.03} & \textbf{1.33}\\
        \hline
    \end{tabular}
    \label{tab:Group_Comp}
\end{table*}

\subsection{Evaluation}
In order to investigate the difference in the performance of the proposed method, we focused on both the individual bone and the whole structure. For evaluating the similarity of the produced segmentation to the ground truth, we utilized the well-known Dice score, which is 
\begin{equation}
    \text{Dice}(y_i,\phi^k(x_i))=\frac{|y_i\cap \phi^k(x_i)|}{(|y_i|+|\phi^k(x_i)|)/2}
\end{equation}, where $y_i$ and $\phi^k(x_i)$ stand for the ground truth and result produced by our segmentation method at $k$-th round. Dice score is able to explain how close two segmentations are, but cannot summarize the accuracy or similarity in a 3-dimension fashion. Thus, we introduced another two metrics for evaluating similarity in 3-D, Average Surface Distance (ASD) and Hausdorff Distance (HD). They can be calculated as follows
\begin{align}
    &\text{ASD}(y_i,\phi^k(x_i))\\
    &=\frac{1}{|y_i|+|\phi^k(x_i)|}(\Sigma_{y \in y_i} d(y,\phi^k(x_i)) + \Sigma_{y'\in \phi^k(x_i)} d(y_i,y')) 
\end{align}

\begin{equation}
        \text{HD}(y_i,\phi^k(x_i))=\text{max}\{d(y_i,\phi^k(x_i)),d(\phi^k(x_i),y_i)\}
\end{equation}, where $d(x,y)$ is the euclidian distance in $\mathbb{R}^2$.

\begin{table*}[ht!]
    \centering
    \caption{Quantitative comparisons between the proposed method and MALF}
    \begin{tabular}{|l|c|c|c|c|c|c|c|c|c|}
        \hline
        \multirow{2}{*}{} & \multicolumn{3}{c|}{\textbf{Humerus}} & \multicolumn{3}{c|}{\textbf{Scapula}} & \multicolumn{3}{c|}{\textbf{Both}}\\
        \cline{2-10}
        & \textit{DSC} & \textit{HD} & \textit{ASD} & \textit{DSC} & \textit{HD} & \textit{ASD} & \textit{DSC} & \textit{HD} & \textit{ASD} \\
        \hline
        MALF & 0.65 & 16.75 & 4.23 & 0.46 & 34.62 & 2.09 & 0.61 & 31.60 & 3.03 \\
        \hline
        \textbf{Our} & \textbf{0.92} & \textbf{4.83} & \textbf{0.72} & \textbf{0.75} & \textbf{21.43} & \textbf{1.27} & \textbf{0.86} & \textbf{20.41} & \textbf{1.02} \\
        \hline
    \end{tabular}
    \label{tab:Mthd_Comp}
\end{table*}
\subsection{Experiment Settings}

One of the main challenges of this task is the high variance between images. Thus normalization among data is required before training. All MR images are cropped and resampled to a voxel spacing of $1mm \times 1mm\times 1mm$  with an image size of $144\times 144\times 80$. Moreover, the N4 bias field correction is used to reduce non-uniformly presented low-frequency noise in the MR images, and the image intensity is mapped into the unit range. Five-fold cross-validation is incorporated in this study to examine the effectiveness of our method. The results are compared to the established atlas segmentation method MALF by using 15 selected images as reference. The proposed network is implemented using Tensorflow framework and the MALF method is carried out with Matlab libraries. The network is trained with one image for each batch, and the Adam solver ($\alpha=0.001$ and learning rate decay by $0.95$ every $10$ epochs) is deployed. 

\subsection{3D CNN Segmentation Models Result}

Table.~\ref{tab:Group_Comp} shows the increase of performance by utilizing our recursive learning framework. The highlighted entries of Table.~\ref{tab:Group_Comp} indicate the performance gain. For dice score, our method can achieve a maximum gain of $20.9\%$ in accuracy for both scapula and humerus segmentation. As for Hausdorff distance and average surface distance, our method reduced the distance between the ground truth and the segmentation result by $67.67\%$ (HD) and $33.36\%$ (ASD) for both bones. As for individual bones, we can see that our method can drastically increase the accuracy of scapula segmentation by over $55\%$ in dice score. By comparing our method to the altas-based registration method that also does not require a large dataset for training in Table.~\ref{tab:Mthd_Comp}, we can see that our method outperforms the MALF method in every metric and with an increase of more than 41\% in dice score. Moreover, the surface distance metric shows that our method has significantly less surface abnormality compared to the traditional registration method.
\subsection{Effectiveness of Recursive Learning}
\begin{table*}[t]
    \centering
    \caption[example]{Evaluation of the effectiveness of recursive learning framework under various dataset sizes}
    \begin{tabular}{|l|c|c|c|c|c|c|c|c|c|}
               \hline
        \multirow{2}{*}{\textbf{\textit{20}}} & \multicolumn{3}{c|}{\textbf{Humerus}} & \multicolumn{3}{c|}{\textbf{Scapula}} & \multicolumn{3}{c|}{\textbf{Both}}\\
        \cline{2-10}
        & \textit{DSC} & \textit{HD} & \textit{ASD} & \textit{DSC} & \textit{HD} & \textit{ASD} & \textit{DSC} & \textit{HD} & \textit{ASD} \\
        \hline
R0 & 0.78 & 22.64 & 3.81 & 0.53 & 34.16 & 3.62 & 0.71 & 34.52 & 2.75\\
\hline
R1 & 0.77 & 18.95 & 2.69 & 0.43 & 36.20 & \textbf{1.22} & 0.69 & \textbf{30.89} & 2.09\\
\hline
R2 & \textbf{0.82} & \textbf{10.29} & \textbf{1.55} & \textbf{0.62} & \textbf{32.49} & 1.27 & \textbf{0.75} & 32.49 & \textbf{1.50}\\
        \hline
        \hline
        \multirow{2}{*}{\textbf{\textit{30}}} & \multicolumn{3}{c|}{\textbf{Humerus}} & \multicolumn{3}{c|}{\textbf{Scapula}} & \multicolumn{3}{c|}{\textbf{Both}}\\
        \cline{2-10}
        & \textit{DSC} & \textit{HD} & \textit{ASD} & \textit{DSC} & \textit{HD} & \textit{ASD} & \textit{DSC} & \textit{HD} & \textit{ASD} \\
        \hline
R0 & 0.80 & 15.60 & 2.42 & 0.51 & 31.78 & 1.66 & 0.71 & 28.85 & 2.17\\
\hline
R1 & 0.81 & 14.05 & 2.06 & 0.59 & 24.78 & 1.56 & 0.74 & 24.61 & 1.80\\
\hline
R2 & \textbf{0.84} & \textbf{10.25} & \textbf{1.71} & \textbf{0.64} & \textbf{19.00} & \textbf{1.42} & \textbf{0.78} & \textbf{18.77} & \textbf{1.53}\\
\hline
        \hline
        \multirow{2}{*}{\textbf{\textit{40}}} & \multicolumn{3}{c|}{\textbf{Humerus}} & \multicolumn{3}{c|}{\textbf{Scapula}} & \multicolumn{3}{c|}{\textbf{Both}}\\
        \cline{2-10}
        & \textit{DSC} & \textit{HD} & \textit{ASD} & \textit{DSC} & \textit{HD} & \textit{ASD} & \textit{DSC} & \textit{HD} & \textit{ASD} \\
        \hline
R0 & 0.80 & 10.68 & 1.46 & 0.52 & 32.29 & 2.08 & 0.71 & 29.70 & 1.69\\
\hline
R1 & 0.83 & 11.45 & 1.48 & \textbf{0.61} & 31.93 & \textbf{1.96} & 0.76 & 27.36 & \textbf{1.62}\\
\hline
R2 & \textbf{0.87} & \textbf{8.27} & \textbf{1.10} & 0.58 & \textbf{28.15} & 2.23 & \textbf{0.78} &\textbf{27.05} & 1.65\\
\hline
\hline
        \multirow{2}{*}{\textbf{\textit{50}}} & \multicolumn{3}{c|}{\textbf{Humerus}} & \multicolumn{3}{c|}{\textbf{Scapula}} & \multicolumn{3}{c|}{\textbf{Both}}\\
        \cline{2-10}
        & \textit{DSC} & \textit{HD} & \textit{ASD} & \textit{DSC} & \textit{HD} & \textit{ASD} & \textit{DSC} & \textit{HD} & \textit{ASD} \\
        \hline
        R0 & 0.78 & 12.87 & 2.16 & 0.39 & 40.14 & 3.32 & 0.66 & 32.80 & 2.20\\
        \hline
        R1 & 0.83 & 12.28 & 1.66 & 0.55 & 26.99 & \textbf{2.26} & 0.75 & 24.02 & 1.79\\
        \hline
        R2 & \textbf{0.88} & \textbf{10.40} & \textbf{1.20} & \textbf{0.61} & \textbf{19.20} & 2.96 & \textbf{0.80} & \textbf{18.74} & \textbf{1.60}\\
        \hline
    \end{tabular}
    \label{tab:Sens_Analys}
\end{table*}

\begin{figure}[ht!]
    \centering
    \includegraphics[width=0.57\linewidth]{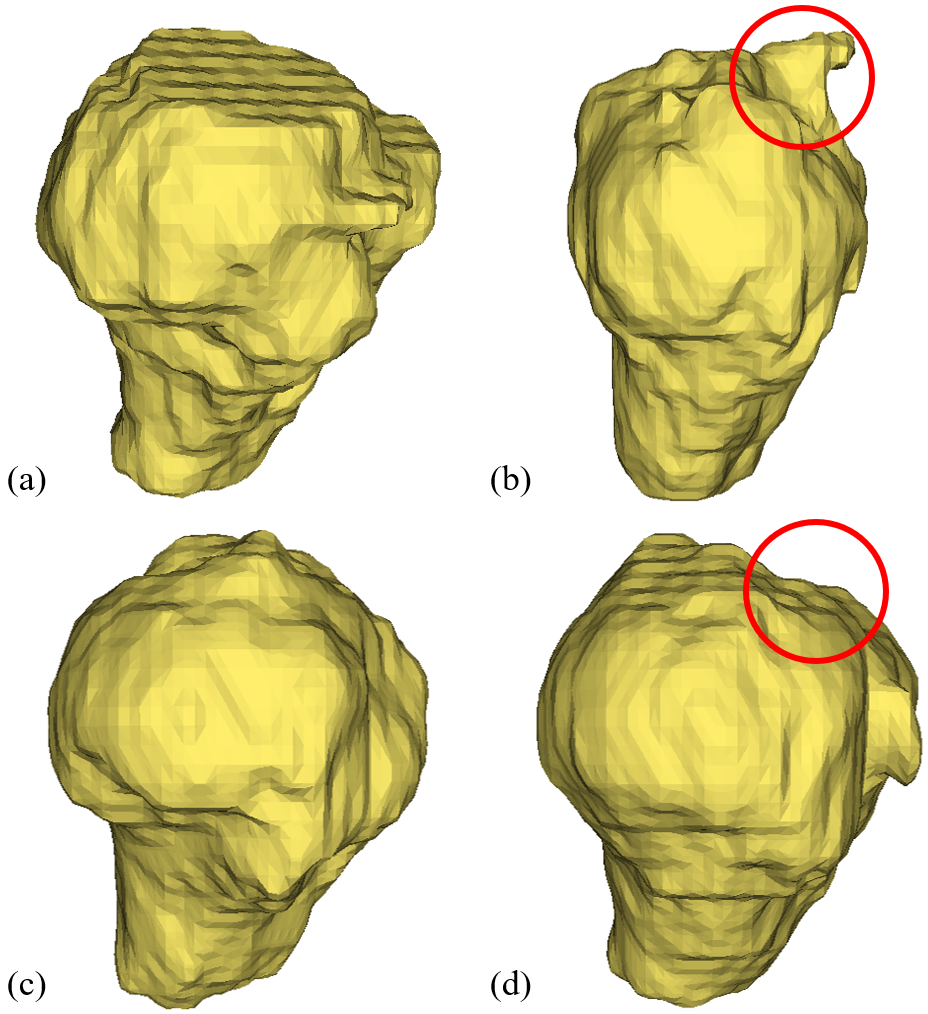}
    \caption{Comparison of humerus between (a) ground truth, (b) segmentation result without Recursive Learning, (c) result after one round of Recursive Learning, (d) result after two rounds of Recursive Learning}
    \label{fig:comp_btw_epoch_humeri}
\end{figure}

\begin{figure}[t]
    \centering
    \includegraphics[width=0.55\linewidth]{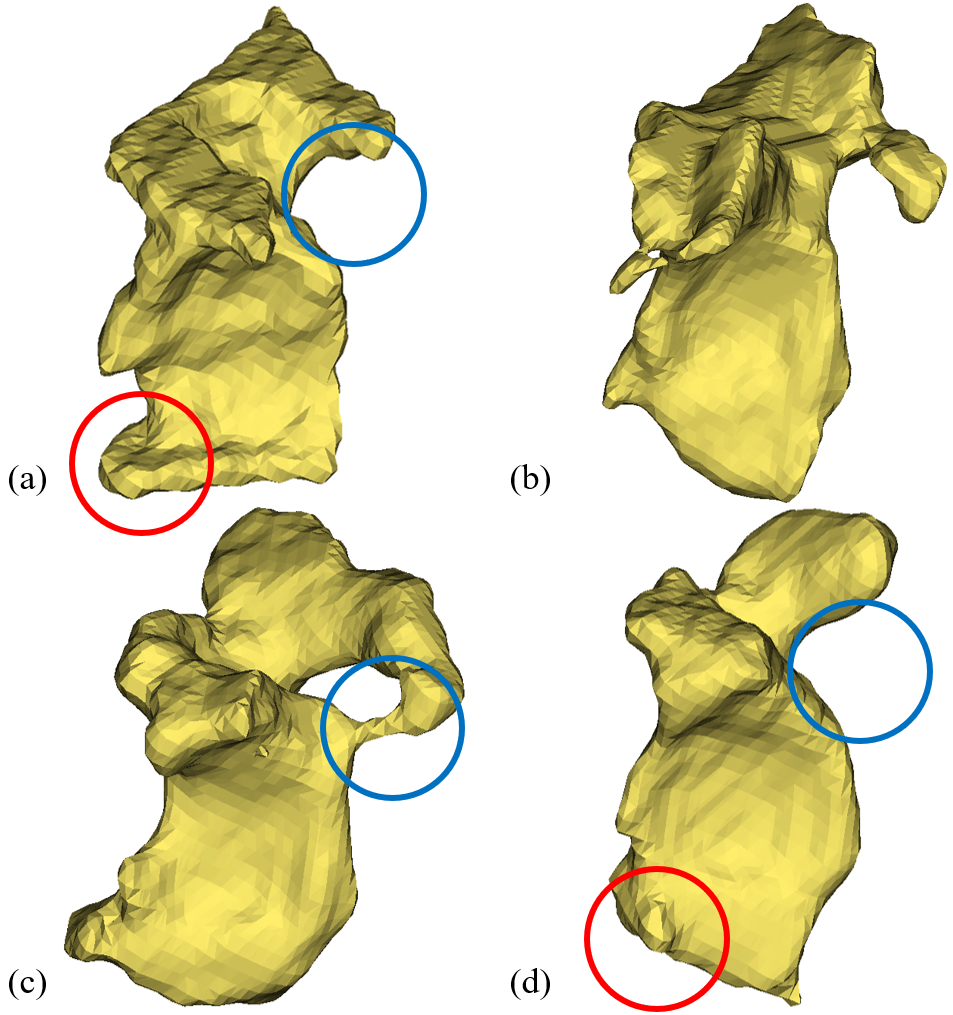}
    \caption{Comparison of glenoid cavity between (a) ground truth, (b) segmentation result without Recursive Learning, (c) result after one round of Recursive Learning, (d) result after two rounds of Recursive Learning}
    \label{fig:sca_gle_c}
\end{figure}

\begin{figure}[t]
    \centering
    \includegraphics[width=0.65\linewidth]{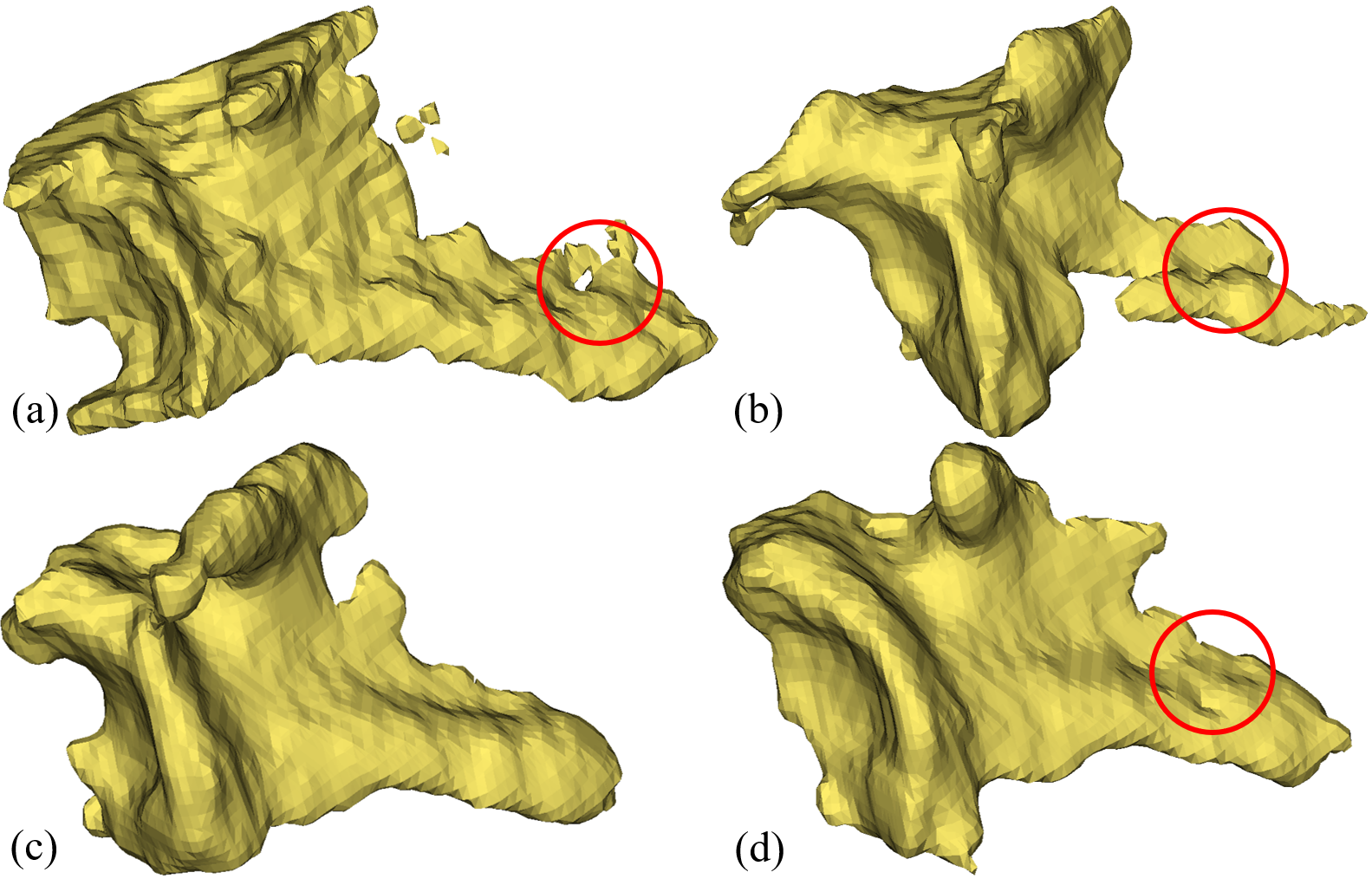}
    \caption{Comparison of subscapula fossa between (a) ground truth, (b) segmentation result without Recursive Learning, (c) result after one round of Recursive Learning, (d) result after two rounds of Recursive Learning}
    \label{fig:sca_fossa}
\end{figure}
In addition to evaluating our proposed method against previously established segmentation algorithms, we performed a further investigation of the effectiveness of the recursive learning framework. To achieve this, we first randomly sampled a reduced amount of data from our 50 image dataset, then we split the training and validation set in $80/20$ and train the proposed network on the training set with the recursive framework. 

Table.~\ref{tab:Sens_Analys} shows that our recursive learning framework helps the proposed network to achieve better performance at an average of $11.7\%$. We tested our method on datasets of sizes ranging from 20 to 50 and performed both similarity and two spatial metrics. However, due to the control of variable, we have only validated 4 images in the size 20 group for evaluating in the distance metric, which have caused the result not to peak at the third round as expected. However, such method shows promising results in dealing with a small dataset with high variance. A more detailed review of the benefits of Recursive Learning will be demonstrated next.

In Fig.~\ref{fig:comp_btw_epoch_humeri}, we demonstrate the 3D surface mesh derived from the ground truth and segmentation results from neural networks with or without recursive learning. All results have undergone the process of eliminating sporadic surfaces and only keeping the largest volume. In part (a) of Fig.~\ref{fig:comp_btw_epoch_humeri}, we can clearly observe the rough surface of the human-produced label, which demonstrated the imperfection of artificial labels. However, such unwanted feature is gradually reduced by our recursive learning framework. Moreover, without the Recursive Learning, the segmentation network lost a majority of the humerus volume and created a non-existing spike indicated in the red circle, which can be perceived by comparing (a) and (b). The spike is removed by training with our proposed method in (c) and (d). In Fig.~\ref{fig:sca_gle_c}, we can see an uneven glenoid cavity as in the red circle due to human labeling along one axis. The glenoid cavity should be smooth and concave in the middle due to the motion between the glenoid and humerus, and the smoothness is progress along with the training rounds. Moreover, focusing on the glenoid head, which is located on the top of Fig.~\ref{fig:sca_gle_c}, the recursive learning framework has the potential to self-correcting errors with the previous segmentation. As in the blue circle in (c) and (d), we can see that the error in (c) is corrected in the same place as (d), thus is more similar to (a). Fig.~\ref{fig:sca_fossa} demonstrate the 3D mesh for subscapular fossa, and clearly, the recursive learning framework reduces the unevenness of the surface and improved the accuracy. Focusing on the red mark in both (a), (b), and (d) for Fig.\ref{fig:sca_fossa}, the discontinuous region exhibited in both ground truth, and the segmentation result with only our deep neural network is reduced in the second round of recursive learning. Therefore, recursive learning can minimize the potential inaccuracy from the ground truth, and previous stages of recursive learning yet yield a more realistic bone structure. In general, as the recursive learning round progress, our recursive learning framework can 1. enhance surface smoothness 2.self-correct errors from previous segmentation 3. reduce human error among ground truth.
\section{Conclusion}
In this paper, we proposed a recursive learning framework together with a deep end-to-end network to segment shoulder joint bones. It is capable of dealing with a small dataset with large parameter variations while improving and reducing errors and human artifacts among ground truth. Various experiments has been carried out to show the effectiveness of completing such task and the benefits gained from deploying recursive learning framework. Besides humerus and scapula, our method can potentially benefit the segmentation of any other bones. 
\printbibliography

\end{document}